\documentclass{article}
\usepackage{arxiv}
\usepackage[utf8]{inputenc} 
\usepackage[T1]{fontenc}    
\usepackage{hyperref}       
\usepackage{url}            
\usepackage{booktabs}       
\usepackage{amsfonts}       
\usepackage{amsmath}
\usepackage{nicefrac}       
\usepackage{microtype}      
\usepackage{cleveref}       
\usepackage{graphicx}
\usepackage{natbib}
\usepackage{doi}
\usepackage[title]{appendix}

\newcommand{\N}{\mathcal{N}}

\title{Estimating ensemble-mean anomaly correlations and the ratio of predictable components using an unbiased estimate of the predictable signal variance}
\author{ \href{https://orcid.org/0000-0002-2958-6637}{\includegraphics[scale=0.06]{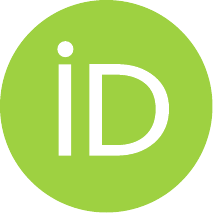}\hspace{1mm}Christopher David Roberts}\\
	ECMWF\\
	Shinfield Park\\
	Reading, United Kingdom\\
	\texttt{chris.roberts@ecmwf.int} \\
}
\renewcommand{\headeright}{Technical note}
\renewcommand{\undertitle}{Technical note}
\renewcommand{\shorttitle}{Estimating ensemble-mean anomaly correlations}

\hypersetup{
  colorlinks   = true, 
  urlcolor     = blue, 
  linkcolor    = blue, 
  citecolor   = red 
}

\begin{document}
\maketitle

\begin{abstract}
The correlation ($r$) between ensemble mean forecasts and observations is a frequently-used metric of forecast quality in the weather and climate prediction communities. However, uncorrected estimates of $r$ are biased towards zero at finite ensemble size due to the influence of unpredictable noise that contaminates the sample ensemble mean. This note summarises established ensemble-size adjustment methods and characterises the behaviour of $r_{\infty}$, a correlation estimator that is constructed from an unbiased and ensemble-size-independent estimate of the signal variance. The only requirement is that forecast members are exchangeable with one another such that they can be interpreted as independent draws from a case-specific predictive distribution. The estimator $r_{\infty}$ does not require forecasts and observations to be statistically exchangeable, only that the unpredictable component of each member is uncorrelated with the observed truth, and thus works for both reliable and unreliable forecasts. In some situations, $r_{\infty}$ may be undefined (i.e. the estimated signal variance is negative) or fall outside the admissible range $-1 \leq r_{\infty} \leq 1$ when the predictable signal is small relative to the sampling uncertainty. These inadmissible values can be interpreted as evidence that the true predictable signal cannot be resolved with the available ensemble size and number of cases. The same unbiased signal variance can also be used to estimate the correlation between the predictable signal and individual members to yield an ensemble-size-invariant estimator of the ratio of predictable components (RPC). In contrast, variance-based RPC estimates, including forms that correct only the model predictable component, are biased low by a factor that depends only on the ensemble size and model signal-to-noise ratio.
\end{abstract}

\section{Relation to previous work}
\label{section:previous}
The signal-plus-noise framework and recovery of infinite-ensemble-size correlation estimates from a finite ensemble are well-established. For example, the adjustment of ensemble-mean correlation skill for ensemble size was described by \citet{murphy1990assessment} and the unbiased estimate of signal variance is a standard result from analysis-of-variance methods \citep[e.g. ][]{rowell1995variability}. Ensemble-size-adjusted correlations are also implemented in the \texttt{EnsCorr} routine of the \texttt{SpecsVerification} package \citep{siegert2017package}. The $r_{\infty}$ estimator described here is equivalent to these earlier results. This note instead emphasises the interpretation of inadmissible estimates as a sampling-uncertainty diagnostic rather than masking them by clipping values to an admissible range, which mirrors the use of physical admissibility constraints to identify statistically paradoxical combinations of reliability metrics that are not possible without sampling uncertainty as described in \citet{roberts2026ensemble}.

The use of ensemble-size-adjusted correlations is also relevant to the ratio of predictable components (RPC) introduced by \citet{eade2014seasonal}, which is commonly estimated in terms of the correlation between ensemble means and observations ($r$) and the correlation between ensemble means and individual members ($r_{mm}$) such that $\textnormal{RPC} = \sqrt { r^2 / r_{mm}^2 } $ \citep{scaife2018signal}. \citet{hardiman2022missing} used the unbiased signal variance to estimate $r_{mm}$, but retained the finite-ensemble correlation between the ensemble mean and observations. Section \ref{section:rpc} shows that applying the unbiased signal variance consistently to both the $r$ and $r_{mm}$ components gives the ensemble-size-invariant estimator $\textnormal{RPC}_{\infty}$. \citet{roberts2026ensemble} obtained an equivalent estimate of $\textnormal{RPC}_{\infty}$ via a different route by ensuring that $r$ and $r_{mm}$ are always calculated in a manner that ensures they remain statistically exchangeable when forecasts are reliable. The $\textnormal{RPC}_{\infty}$ presented here instead adjusts each correlation separately using the unbiased signal variance.

\section{Notation and anomaly definition}
\label{section:notation}
Consider an ensemble forecast system with $k=1,\dots,N$ members covering $j=1,\dots,M$ independent cases (e.g. forecast start dates) such that ensemble forecast members and the observed truth are denoted using $x_{1,j},\dots,x_{N,j}$ and $x_{T,j}$, respectively. We define ensemble forecast anomalies ($z_{k,j}$) and observed anomalies ($z_{T,j}$) relative to `by-member--all-years' climatologies such that

\begin{equation}
      z_{k,j} = x_{k,j} - \frac{1}{L} \sum\limits^{L}_{h=1} x_{k,h},
\end{equation}
\begin{equation}
      z_{T,j} = x_{T,j} - \frac{1}{L} \sum\limits^{L}_{h=1} x_{T,h},
\end{equation}

where $L$ is the number of years in the forecast dataset and $h=1,\dots,L$ represents the subset of all cases with the same calendar start date as case $j$. Anomalies are thus calculated relative to climatologies estimated separately for each member and each start date. This approach has no impact on ensemble means, but ensures that forecast member anomalies remain statistically exchangeable with observed anomalies if the underlying raw forecasts are perfectly reliable. See \citet{roberts2025unbiased} for further details.

\section{Unadjusted ensemble mean anomaly correlations}
\label{section:correlations}
For finite ensemble size, the uncorrected anomaly correlation between the observed truth and ensemble mean forecast anomaly is given by

\begin{equation}
      \label{equation:correlation}
      r= \frac{\mathbb{E}\left[ \left<z_{.,j}\right>_{N}  z_{T,j}  \right]}{\sqrt{\mathbb{E}\left[ \left<z_{.,j}\right>_{N} ^2 \right]   \mathbb{E}\left[ z_{T,j}^2  \right]}},
\end{equation}
where $\mathbb{E} \left[ \cdot \right]$ is the expectation over cases $j$, $\mathbb{E}\left[z_{T,j}\right] = \mathbb{E}\left[z_{k,j}\right] = 0$, and $\left<\cdot\right>_N$ represents the mean over a sample of $N$ members such that the ensemble mean for case $j$ is denoted $\left<z_{.,j}\right>_N \equiv \frac{1}{N}\sum^{N}_{k=1}z_{k,j}$. For a finite ensemble, $\mathbb{E}\left[ \left<z_{.,j}\right>_N^2 \right]$ is biased high compared to the variance of the true predictable signal such that estimates of $r$ are biased towards zero.

\section{Adjusted ensemble mean anomaly correlations}
\label{section:adjusted_correlations}
If forecast member anomalies can be interpreted as conditionally independent\footnote{i.e.\ independent of one another given the shared predictable signal.} draws from a common predictive distribution at each case $j$ they can be decomposed as

\begin{equation}
    \label{equation:decomposition}
    z_{k,j} = s_j + n_{k,j},
\end{equation}
where $s_j$ is the true predictable signal (i.e. $\left<z_{.,j}\right>_N \to s_j$ as $N \to \infty$) and $n_{k,j}$ are independent realisations of unpredictable noise such that $\mathbb{E}\left[ n_{k,j} s_{j} \right]=0$ and $\mathbb{E}\left[ n_{k,j} n_{l,j} \right]=0$ (where $k \neq l$). From these properties it follows that

\begin{equation}
    \label{equation:numerator}
    \mathbb{E}\left[ \left<z_{.,j}\right>_N z_{T,j} \right] = \frac{1}{N}\sum^{N}_{k=1} \mathbb{E}\left[ z_{k,j} z_{T,j} \right] = \mathbb{E}\left[ z_{k,j} z_{T,j} \right] = \mathbb{E}\left[ s_{j} z_{T,j} \right],
\end{equation}
which is independent of ensemble size and equal to the covariance between the predictable signal and the observed truth. The only assumption is that $\mathbb{E}\left[ n_{k,j} z_{T,j} \right]=0$ such that the unpredictable component of each forecast member is uncorrelated with the observed truth. There is no requirement for the forecast to be reliable or to share the same predictable signal as the truth. The variance of the ensemble mean anomaly can then be expanded as a double sum and separated into $N$ diagonal terms ($k=l$) and $N(N-1)$ off-diagonal terms ($k \neq l$) to give
\begin{equation}
    \label{equation:denominator}
    \begin{split}
        \mathbb{E}\left[ \left<z_{.,j}\right>_N^2 \right] &= \frac{1}{N^2}\sum^{N}_{k=1}\sum^{N}_{l=1} \mathbb{E}\left[ z_{k,j} z_{l,j} \right] \\
        &= \frac{1}{N^2}\left( N\,\mathbb{E}\left[ z_{k,j}^2 \right] + N(N-1)\,\mathbb{E}\left[ z_{k,j} z_{l,j} \right]_{k \neq l} \right) \\
        &= \frac{1}{N}\mathbb{E}\left[ z_{k,j}^2 \right] + \frac{N-1}{N}\mathbb{E}\left[ z_{k,j} z_{l,j} \right]_{k \neq l}.
    \end{split}
\end{equation}

Using the properties $\mathbb{E}\left[ n_{k,j} s_{j} \right]=0$ and $\mathbb{E}\left[ n_{k,j} n_{l,j} \right]=0$ ($k \neq l$) introduced with equation \ref{equation:decomposition}, the off-diagonal component can be rewritten as

\begin{equation}
    \label{equation:between_member}
    \mathbb{E}\left[ z_{k,j} z_{l,j} \right]_{k \neq l} = \mathbb{E}\left[ s_j^2 \right],
\end{equation}
where $\mathbb{E}\left[ s_j^2 \right]$ is the variance of the true predictable signal that is independent of ensemble size. Substitution of equation \ref{equation:between_member} into equation \ref{equation:denominator} then gives the following unbiased estimator of the true signal variance

\begin{equation}
    \label{equation:unbiased_var}
    \mathbb{E}\left[ s_j^2 \right] = \frac{N\,\mathbb{E}\left[ \left<z_{.,j}\right>_N^2 \right] - \mathbb{E}\left[ z_{k,j}^2 \right]}{N-1}.
\end{equation}

Equation \ref{equation:unbiased_var} can be rearranged into the algebraically equivalent, and perhaps more intuitive, form
\begin{equation}
    \label{equation:unbiased_var_alt}
    \mathbb{E}\left[ s_j^2 \right] = \mathbb{E}\left[ \left<z_{.,j}\right>_N^2 \right] - \frac{1}{N}\,\mathbb{E}\left[ \hat{\sigma}^2_{n,j} \right],
\end{equation}
where
\begin{equation}
    \label{equation:ensemble_variance}
    \hat{\sigma}^2_{n,j} = \frac{1}{N-1}\sum^{N}_{k=1}\left( z_{k,j} - \left<z_{.,j}\right>_N \right)^2
\end{equation}
is the unbiased variance of ensemble perturbations about the sample ensemble mean for case $j$, which is an estimate of the unpredictable noise variance $\mathbb{E}\left[ n_{k,j}^2 \right]$. Substitution of either \ref{equation:unbiased_var} or  \ref{equation:unbiased_var_alt} into equation \ref{equation:correlation} yields the following estimator for the ensemble mean anomaly correlation
\begin{equation}
    \label{equation:unbiased_correlation}
    r_{\infty} = \frac{\mathbb{E}\left[ \left<z_{.,j}\right>_N z_{T,j} \right]}{\sqrt{\mathbb{E}\left[ s_j^2 \right] \mathbb{E}\left[ z_{T,j}^2 \right]}},
\end{equation}
where the use of $\mathbb{E}\left[ s_j^2 \right]$ ensures that $r_{\infty}$ targets the correlation between the predictable signal and the observed truth ($r_{\infty} \to \textnormal{Corr}(s_j, z_{T,j})$ as $M \to \infty$) and is free from systematic ensemble-size bias for both reliable and unreliable forecasts. A small residual bias remains at finite $M$ because $r_{\infty}$ is a nonlinear function of sample estimates, but this vanishes as $M \to \infty$.

\section{Sampling effects}
\label{section:caveats}

The signal variance estimates given by equations \ref{equation:unbiased_var} and \ref{equation:unbiased_var_alt} are unbiased in expectation, but finite sample estimates may be physically or statistically inadmissible when the predictable signal is small relative to the sampling uncertainty. Firstly, negative estimates of $\mathbb{E}\left[ s_j^2 \right]$ are possible and should be interpreted as evidence that the predictable signal is statistically indistinguishable from zero at the available sample size. For example, in the case that $\mathbb{E}\left[ s_j^2 \right]=0$ an unbiased estimator will sometimes return values that are negative. However, even when the estimated signal variance is positive, sampling variability can yield  $r_{\infty}$ estimates outside the admissible range $-1 \leq r_{\infty} \leq 1$, which cannot be interpreted as correlations. Previous studies \citep[e.g. ][]{rowell1995variability} and the \texttt{SpecsVerification} package \citep{siegert2017package} handle these edge cases by clipping such that the adjusted correlations remain within the admissible range. However, this note argues that these inadmissible values should not be masked but instead reported and interpreted as evidence that the true predictable signal cannot be resolved at the available ensemble size $N$ and number of cases $M$.

There is also a subtlety relating to the expectation-based correlation definitions used in equations \ref{equation:correlation} and \ref{equation:unbiased_correlation}, which are only valid under the implicit assumption that the anomalies are zero-mean, i.e. $\mathbb{E}\left[ z_{k,j} \right] = \mathbb{E}\left[ z_{T,j} \right] = 0$. This assumption holds by construction when anomalies are defined relative to a climatology estimated over exactly the same set of cases $j$ used to evaluate the correlation. However, this can be broken if anomalies are first computed over a full dataset and then correlations evaluated over subsets of cases for which the anomalies have a non-zero sample mean. This issue can be avoided by recomputing anomalies for subset of cases over which the correlation is evaluated.

\section{Implications for the ratio of predictable components (RPC)}
\label{section:rpc}
The RPC compares the correlation between the ensemble mean and the observed truth ($r$) with the correlation between the ensemble mean and an individual forecast member ($r_{mm}$). RPC values greater than one correspond to the counterintuitive situation where the real world appears to be more predictable than individual ensemble members from the same forecast model. When combined with observations and forecast members that have the same total variance, this type of unreliability has been described as a `signal-to-noise paradox' \citep[SNP; ][]{eade2014seasonal, scaife2018signal}. Each RPC estimator below is labelled by the initials of the authors and the year of the study that introduced it. The original variance-based RPC definition of \citet{eade2014seasonal} used estimates of the model predictable component derived from the variance of the ensemble mean, such that
\begin{equation}
    \label{equation:rpc_var}
    \textnormal{RPC}_{\textnormal{E14}} = \frac{|r|}{\sqrt{\mathbb{E}\left[ \left<z_{.,j}\right>_N^2 \right] / \mathbb{E}\left[ z_{k,j}^2 \right]}}.
\end{equation}
In this case, $\mathbb{E}\left[ \left<z_{.,j}\right>_N^2 \right]$ overestimates the variance of the predictable signal and $|r|$ is biased low, such that $\textnormal{RPC}_{\textnormal{E14}}$ is biased low at finite $N$. The magnitude of $r$ is used here, and in equation \ref{equation:rpc_h}, for consistency with the correlation-based definition introduced in equation \ref{equation:rpc_corr_trad}. \citet{hardiman2022missing} replaced the ensemble mean variance in the denominator of equation \ref{equation:rpc_var} with the unbiased estimate of the signal variance (equation \ref{equation:unbiased_var}), such that
\begin{equation}
    \label{equation:rpc_h}
    \textnormal{RPC}_{\textnormal{H22}} = \frac{|r|}{\sqrt{\mathbb{E}\left[ s_j^2 \right] / \mathbb{E}\left[ z_{k,j}^2 \right]}}.
\end{equation}
This removes the ensemble-size dependence of the model predictable component, but the numerator retains the finite-ensemble correlation $r$ such that the resulting RPC estimates are still biased low for finite $N$. The correlation-based definition of RPC \citep{scaife2018signal} instead compares the model-observation and model-model correlations directly, such that
\begin{equation}
    \label{equation:rpc_corr_trad}
    \textnormal{RPC}_{\textnormal{SS18}} = \sqrt{\frac{r^2}{r_{mm}^2}},
\end{equation}
where $r$ is the correlation between the observed truth and the ensemble mean (equation \ref{equation:correlation}) and $r_{mm}$ is the mean over members $k$ of the correlation between member $k$ and the mean of the remaining members. Both correlations are attenuated by the unpredictable noise in a finite ensemble mean. If $r$ is calculated using all $N$ members while $r_{mm}$ uses ensemble means of $N-1$ members, the numerator is attenuated less than the denominator and the resulting estimate is biased high at finite $N$.

\citet{roberts2026ensemble} formalised an ensemble-size-unbiased version of equation \ref{equation:rpc_corr_trad} in which the correlation between the observed truth and the ensemble mean is calculated using only $N-1$ members, such that
\begin{equation}
    \label{equation:r_rv26}
    r_{N-1} = \frac{\mathbb{E}\left[ \left<z_{.,j}\right>_{N-1} z_{T,j} \right]}{\sqrt{\mathbb{E}\left[ \left<z_{.,j}\right>_{N-1}^2 \right] \mathbb{E}\left[ z_{T,j}^2 \right]}},
\end{equation}
where $\left<\cdot\right>_{N-1}$ denotes the mean over the first $N-1$ members. Each model-model correlation is calculated between member $k$ and the mean of the remaining $N-1$ members, such that
\begin{equation}
    \label{equation:rmm_k}
    r_{mm(k)} = \frac{\mathbb{E}\left[ \left<z_{.,j}\right>_{N-1}^{i \neq k} z_{k,j} \right]}{\sqrt{\mathbb{E}\left[ \left( \left<z_{.,j}\right>_{N-1}^{i \neq k} \right)^2 \right] \mathbb{E}\left[ z_{k,j}^2 \right]}},
\end{equation}
where $\left<z_{.,j}\right>_{N-1}^{i \neq k} \equiv \frac{1}{N-1}\sum^{N}_{\substack{i=1\\i \neq k}} z_{i,j}$, so the excluded member never contributes to the ensemble mean it is compared with. These correlations are combined as
\begin{equation}
    \label{equation:rpc_rv26}
    \textnormal{RPC}_{\textnormal{RV26}} = \sqrt{\frac{r_{N-1}^2}{r_{mm}^2}}, \qquad r_{mm} = \frac{1}{N}\sum^{N}_{k=1} r_{mm(k)}.
\end{equation}
The model-observation and model-model correlations are then both calculated from ensemble means of the same size, such that they are statistically exchangeable when the forecast is reliable. More generally, the unpredictable noise in the ensemble mean attenuates both correlations by the same factor, regardless of reliability, such that $\textnormal{RPC}_{\textnormal{RV26}}$ is independent of ensemble size. See \citet{roberts2026ensemble} for further details.

An alternative route to the same result is to estimate all correlations using the procedure described in section \ref{section:adjusted_correlations}. Starting from equation \ref{equation:decomposition} and using $\mathbb{E}\left[ n_{k,j} s_j \right]=0$, the covariance between the model predictable signal and an individual member is $\mathbb{E}\left[ s_j z_{k,j} \right] = \mathbb{E}\left[ s_j^2 \right]$. The correlation between the predictable signal and an individual member is therefore
\begin{equation}
    \label{equation:rmm_inf}
    r_{mm,\infty} = \textnormal{Corr}(s_j, z_{k,j}) = \frac{\mathbb{E}\left[ s_j z_{k,j} \right]}{\sqrt{\mathbb{E}\left[ s_j^2 \right] \mathbb{E}\left[ z_{k,j}^2 \right]}} = \sqrt{\frac{\mathbb{E}\left[ s_j^2 \right]}{\mathbb{E}\left[ z_{k,j}^2 \right]}},
\end{equation}
which is the model-model correlation that would be achieved by an infinite ensemble. When $\mathbb{E}\left[ s_j^2 \right]$ is estimated using equation \ref{equation:unbiased_var}, $r_{mm,\infty}$ is identical to the denominator of $\textnormal{RPC}_{\textnormal{H22}}$ from equation \ref{equation:rpc_h} as derived by \citet{hardiman2022missing}. Combining $r_{\infty}$ with $r_{mm,\infty}$ then gives
\begin{equation}
    \label{equation:rpc_inf}
    \textnormal{RPC}_{\infty} = \frac{|r_{\infty}|}{r_{mm,\infty}} = \frac{\left|\mathbb{E}\left[ \left<z_{.,j}\right>_N z_{T,j} \right]\right| \sqrt{\mathbb{E}\left[ z_{k,j}^2 \right]}}{\mathbb{E}\left[ s_j^2 \right] \sqrt{\mathbb{E}\left[ z_{T,j}^2 \right]}},
\end{equation}
which provides an ensemble-size-invariant estimate of RPC. This is the same quantity targeted by $\textnormal{RPC}_{\textnormal{RV26}}$ by a different route. Comparison of equation \ref{equation:rpc_inf} with equations \ref{equation:rpc_var} and \ref{equation:rpc_h} gives
\begin{equation}
    \label{equation:rpc_bias}
    \textnormal{RPC}_{\textnormal{E14}} = \gamma_N \, \textnormal{RPC}_{\infty}, \qquad \textnormal{RPC}_{\textnormal{H22}} = \sqrt{\gamma_N} \, \textnormal{RPC}_{\infty},
\end{equation}
where
\begin{equation}
    \label{equation:gamma}
    \gamma_N = \frac{\mathbb{E}\left[ s_j^2 \right]}{\mathbb{E}\left[ \left<z_{.,j}\right>_N^2 \right]} = \frac{N r_{mm,\infty}^2}{1 + (N-1)\, r_{mm,\infty}^2}
\end{equation}
follows from rearranging equation \ref{equation:unbiased_var} for $\mathbb{E}\left[ \left<z_{.,j}\right>_N^2 \right]$ and using equation \ref{equation:rmm_inf}. Both variance-based estimates are therefore biased low at finite $N$ by a factor that depends only on $N$ and the model signal-to-noise ratio. $\textnormal{RPC}_{\infty}$ uses all $N$ members in the ensemble mean and requires no leave-one-out calculations. However, it is undefined when the estimated signal variance is non-positive and inadmissible when $|r_{\infty}| > 1$. In contrast, $\textnormal{RPC}_{\textnormal{RV26}}$ is always defined.

\section{Examples using idealized Gaussian forecasts}
\label{section:examples}
This section illustrates the differences between the biased and unbiased sample correlation estimators described above using idealised data. Let $\{x_{k,j}\}$ represent an ensemble forecast of \(k=1,\dots,N\) members covering \(j=1,\dots,M\) independent cases with $\{y_j\}$ verifying observations, generated by a signal-plus-noise model

\begin{align}
    \label{eq:gaussian_data}
    x_{k,j} &= s_j + n_{k,j}, \\
    y_j &= s_j + e_j,
\end{align}
where $s_j$, $n_{k,j}$, and $e_j$ are mutually independent with $s_j \sim \N(0, \sigma_s^2)$, $n_{k,j} \sim \N(0, \alpha^2)$, and $e_j \sim \N(0, \beta^2)$. These forecasts are unbiased but potentially unreliable. The true correlation between the predictable signal and the observations is $\rho = \textnormal{Corr}(s_j, y_j) = \sigma_s / \sqrt{\sigma_s^2 + \beta^2}$. In the examples below, the sample estimators $r$ (equation \ref{equation:correlation}) and  $r_{\infty}$ (equation \ref{equation:unbiased_correlation}) are shown as a function of ensemble size $N$, with solid lines and shaded bands indicating the median and 10th--90th percentile range over 1000 independent realisations, respectively. Grey lines show the percentage of inadmissible estimates of $r_{\infty}$, which should be considered a measure of the uncertainty in the true predictable signal as described in section \ref{section:caveats}.

\begin{itemize}
    \item \textbf{Reliable, large sample, strong signal}. In this reliable case, $r$ is biased low for small $N$ and converges towards $\rho$ as $N \to \infty$. In contrast, $r_{\infty}$ shows no systematic dependence on $N$ and is centred on $\rho$.
    \begin{center}
        \includegraphics[width=0.68\linewidth]{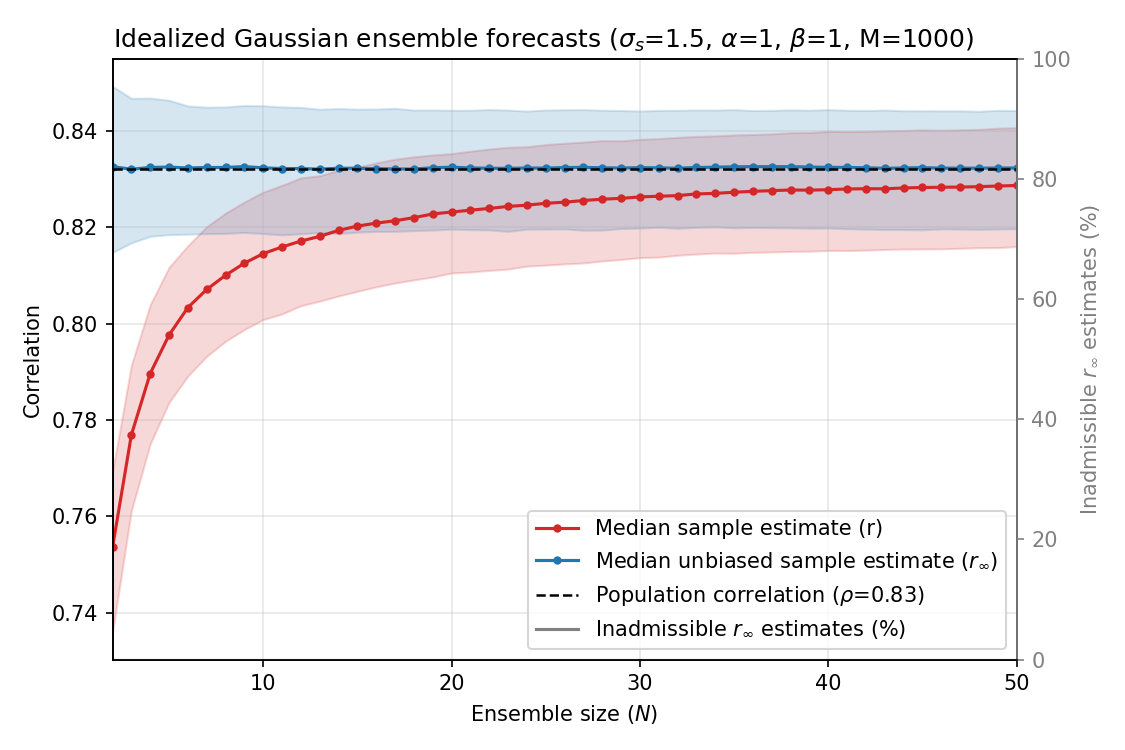}
    \end{center}

    \item \textbf{Overdispersive, large sample, strong signal}. In this overdispersive case, the increased noise of forecast members increases the low bias of $r$, but $r_{\infty}$ still recovers $\rho$ on average.
    \begin{center}
        \includegraphics[width=0.68\linewidth]{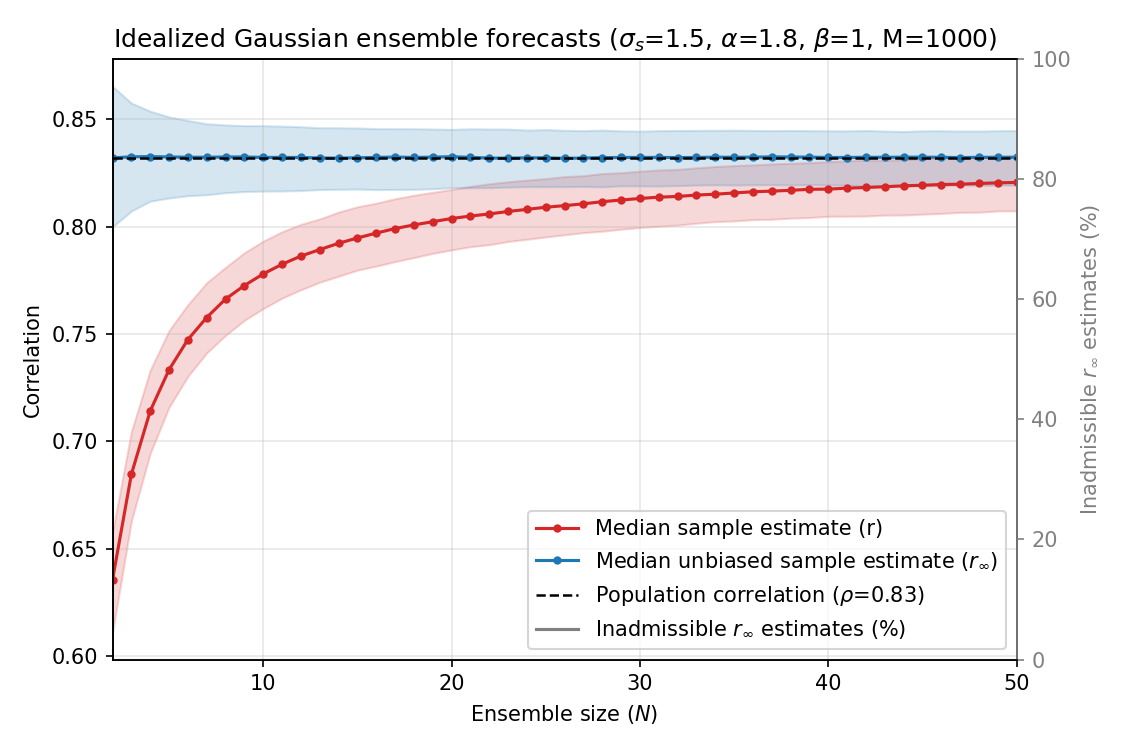}
    \end{center}

    \item \textbf{Underdispersive, large sample, strong signal}. In this underdispersive case, the reduced noise of forecast members reduces the low bias of $r$, but $r_{\infty}$ again recovers $\rho$ on average.
    \begin{center}
        \includegraphics[width=0.68\linewidth]{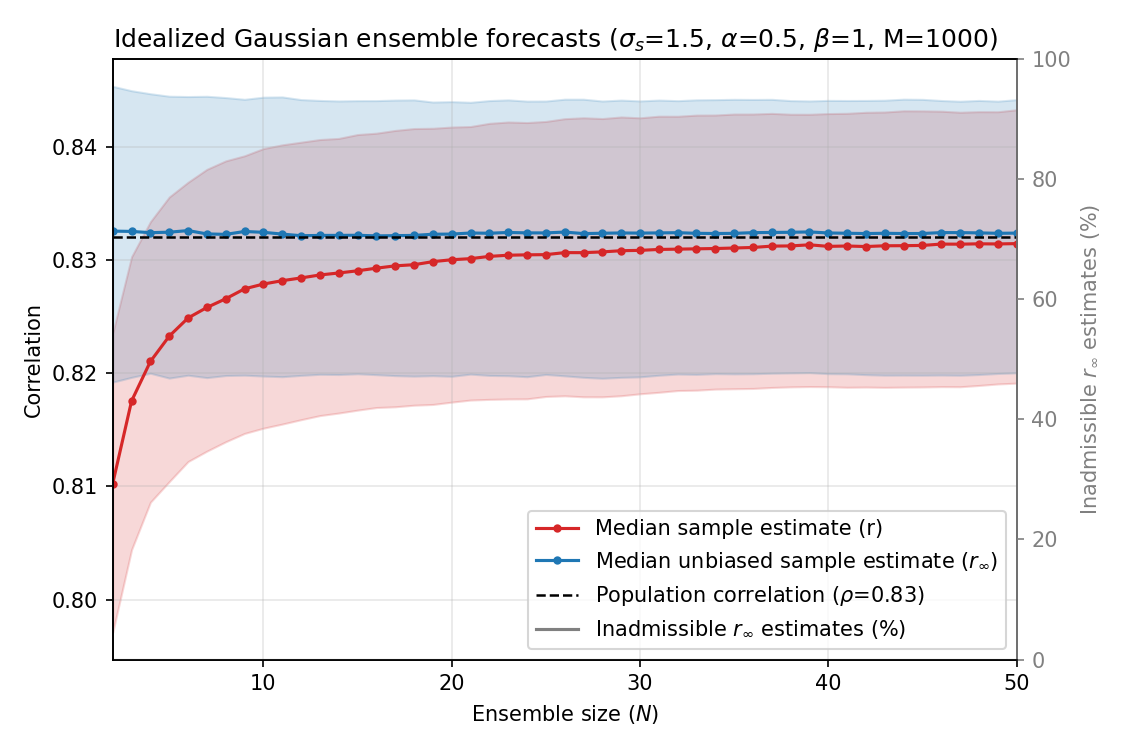}
    \end{center}

    \item \textbf{Reliable, large sample, weak signal}. In this reliable example, the signal is weak but the large sample keeps the sampling uncertainty small, so $r_{\infty}$ recovers $\rho$ on average.
    \begin{center}
        \includegraphics[width=0.68\linewidth]{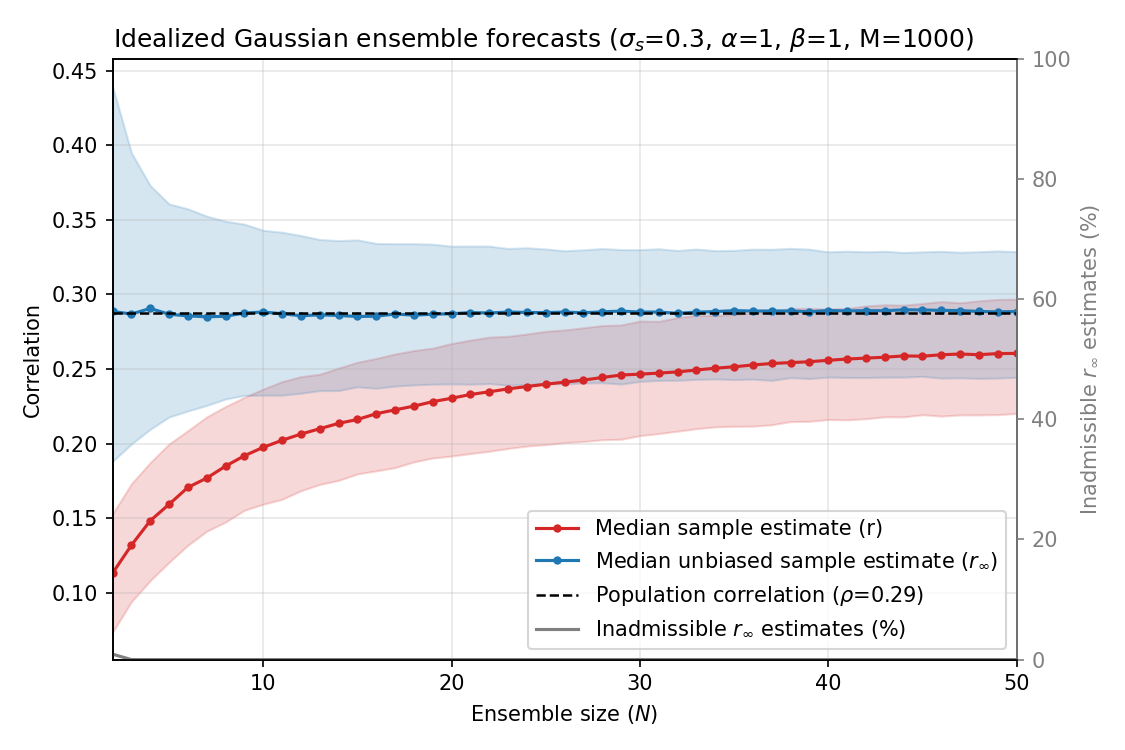}
    \end{center}

    \item \textbf{Overdispersive, large sample, weak signal}. In this overdispersive example, the increased noise of forecast members further increases the low bias of $r$, but the large sample still allows $r_{\infty}$ to recover $\rho$ on average. For very small ensemble sizes ($N < 5$) sampling uncertainties become very large and there are some inadmissible $r_{\infty}$ estimates (see section \ref{section:caveats}).
    \begin{center}
        \includegraphics[width=0.68\linewidth]{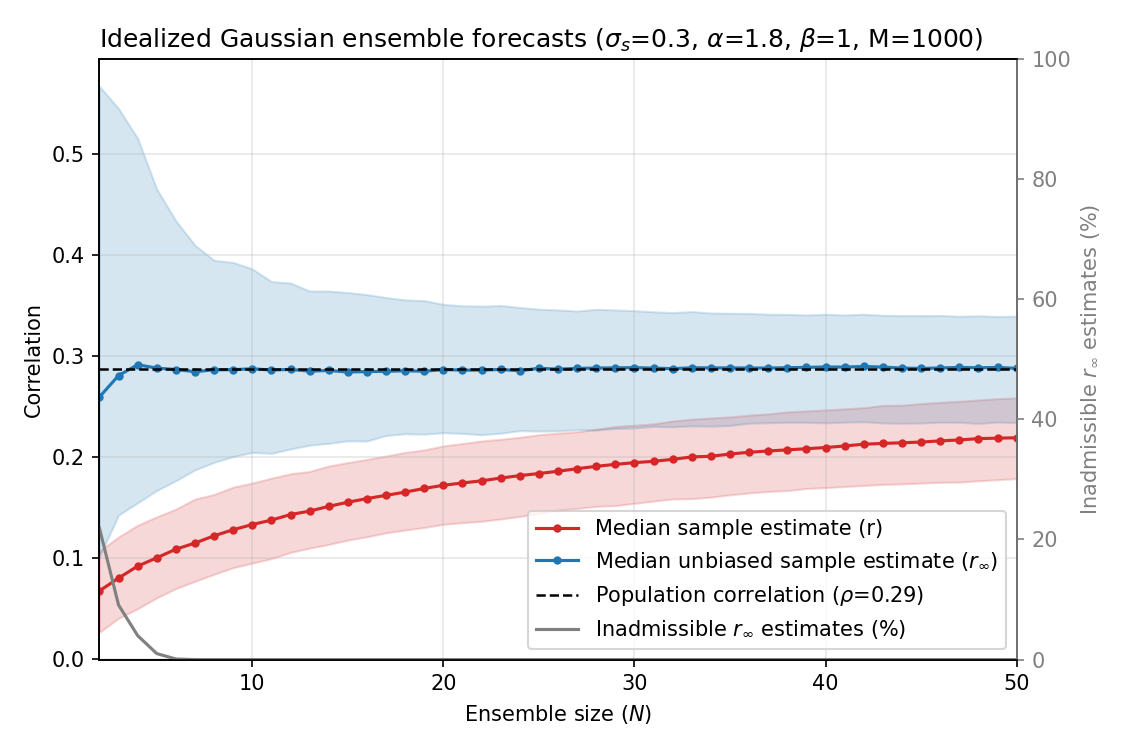}
    \end{center}

    \item \textbf{Reliable, small sample, weak signal}. In this reliable example, the weak signal and small $M$ result in large sampling uncertainties and $r_{\infty}$ is frequently undefined or falls outside the range $[-1, 1]$.
    \begin{center}
        \includegraphics[width=0.68\linewidth]{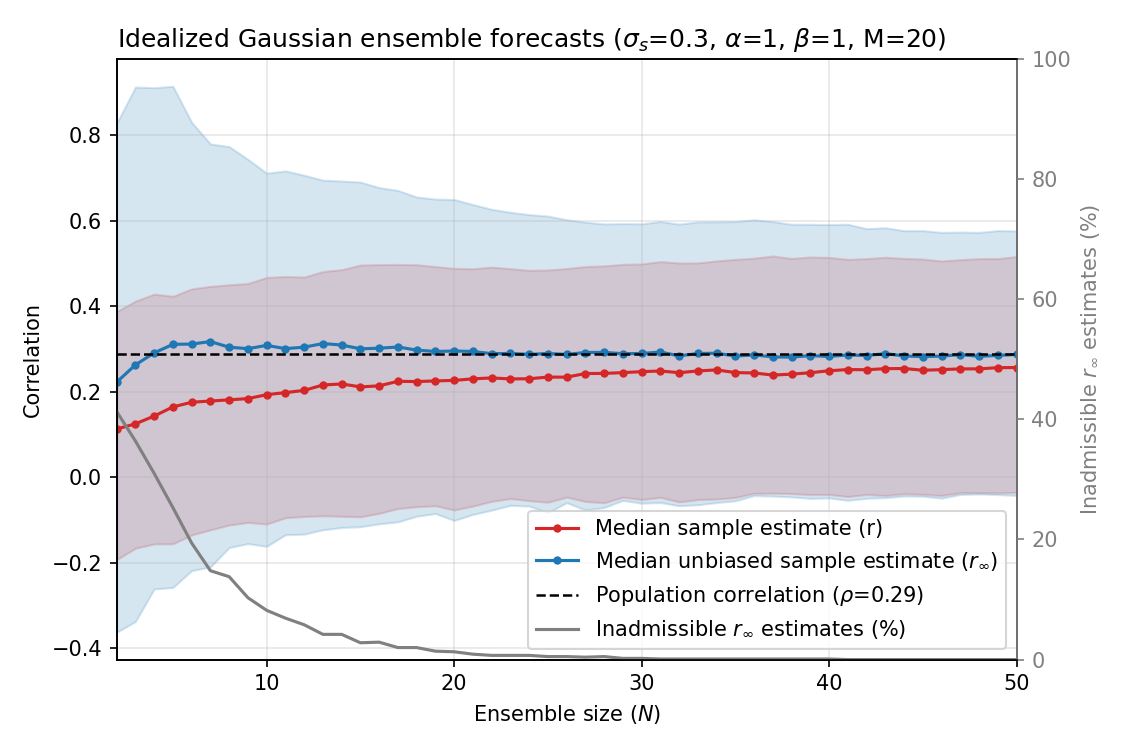}
    \end{center}

    \item \textbf{Overdispersive, small sample, weak signal}. In this overdispersive example, the increased noise of forecast members combined with a weak signal and large sampling uncertainties (i.e. small $M$) substantially increases the chance of physically inadmissible estimates of $r_{\infty}$.
    \begin{center}
        \includegraphics[width=0.68\linewidth]{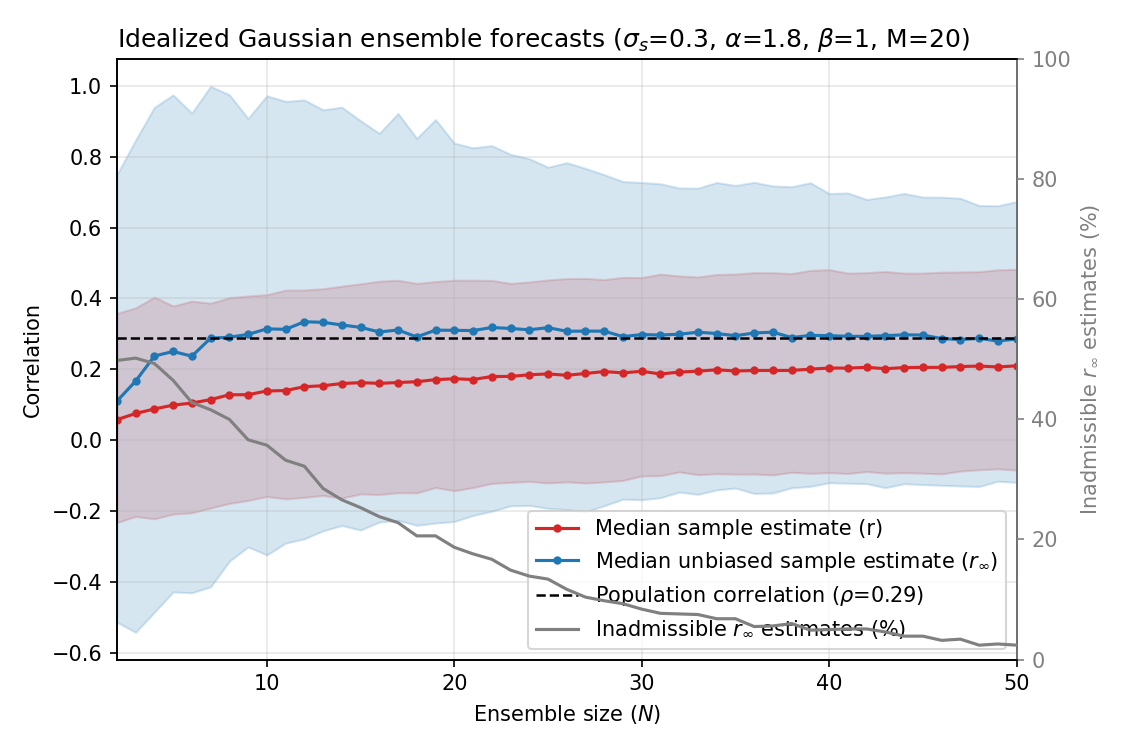}
    \end{center}
\end{itemize}

The same idealised data are used below to compare $\textnormal{RPC}_{\textnormal{E14}}$, $\textnormal{RPC}_{\textnormal{H22}}$, $\textnormal{RPC}_{\textnormal{RV26}}$, and $\textnormal{RPC}_{\infty}$ (section \ref{section:rpc}). The population RPC is $\sqrt{\sigma_s^2 + \alpha^2} / \sqrt{\sigma_s^2 + \beta^2}$, which equals one for a reliable forecast. Dotted lines show the expected values of $\textnormal{RPC}_{\textnormal{E14}}$ and $\textnormal{RPC}_{\textnormal{H22}}$ from equation \ref{equation:rpc_bias}. Medians and percentiles of $\textnormal{RPC}_{\textnormal{H22}}$ and $\textnormal{RPC}_{\infty}$ are computed only over realisations with a positive signal variance estimate.

\begin{itemize}
    \item \textbf{Overdispersive, large sample, weak signal}. This is example of an unreliable forecast with a population RPC greater than one. Both variance-based estimates are biased low and follow the curves predicted by equation \ref{equation:rpc_bias}, so that $\textnormal{RPC}_{\textnormal{E14}}$ and $\textnormal{RPC}_{\textnormal{H22}}$ remain substantially below the population RPC even for $N=50$. $\textnormal{RPC}_{\textnormal{RV26}}$ and $\textnormal{RPC}_{\infty}$ are centred on the population value and have very similar sampling distributions.
    \begin{center}
        \includegraphics[width=0.68\linewidth]{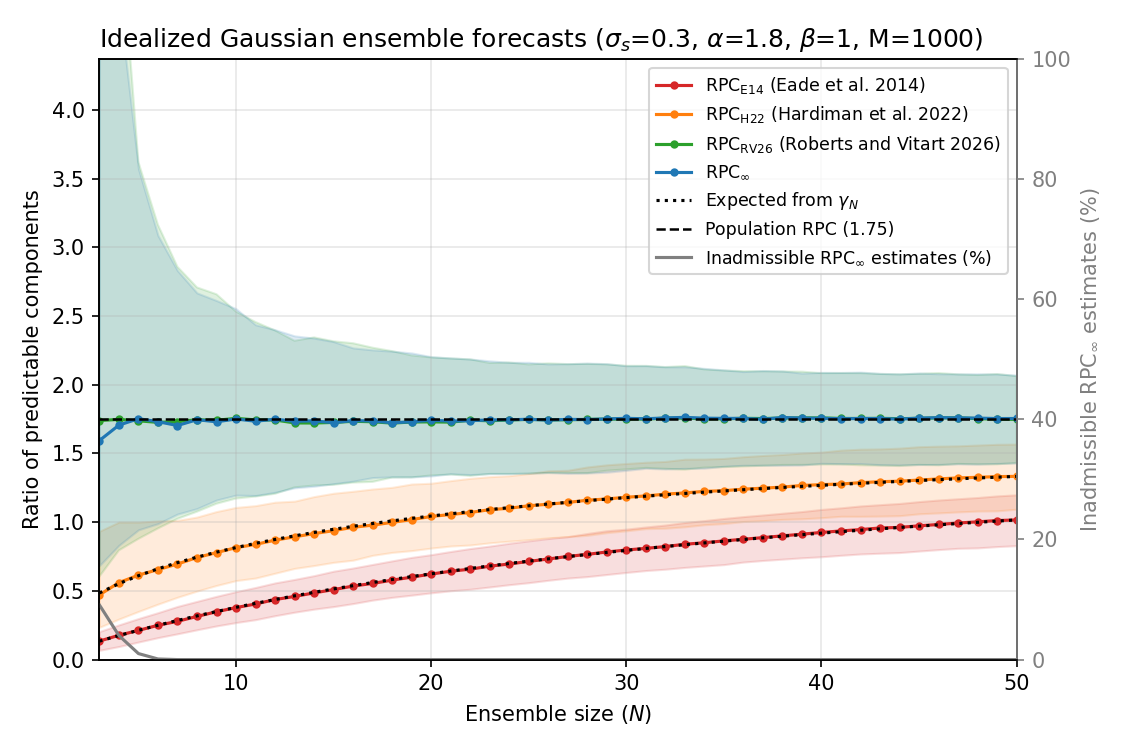}
    \end{center}

    \item \textbf{Underdispersive, large sample, weak signal}. This is an example of an unreliable forecast with a population RPC that is less than one. The variance-based estimates are again biased low, while $\textnormal{RPC}_{\textnormal{RV26}}$ and $\textnormal{RPC}_{\infty}$ show no systematic dependence on $N$.
    \begin{center}
        \includegraphics[width=0.68\linewidth]{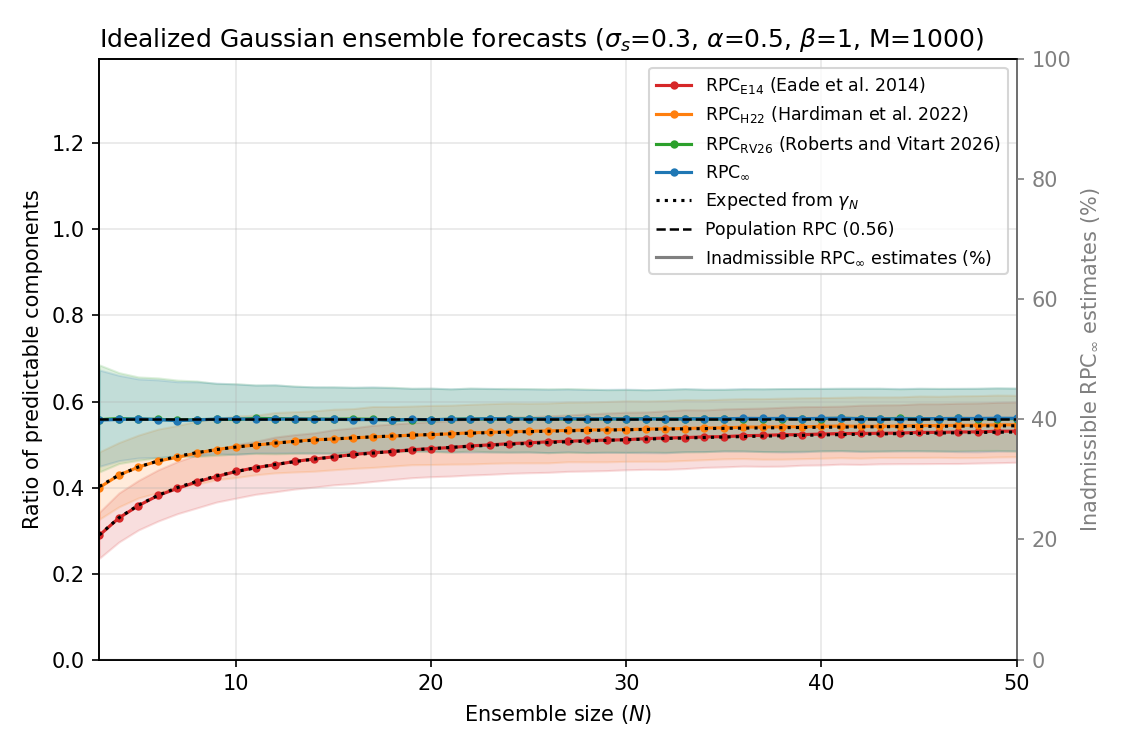}
    \end{center}
\end{itemize}

\section{Summary}
\label{section:summary}
The purpose of this note was to bring together and summarise existing results rather than to derive new methods. The signal-plus-noise decomposition, the unbiased estimate of signal variance, and ensemble-size adjustments of correlation skill are all well established, and the results presented here follow from these using elementary algebra. The aim was to state these results in a consistent notation, clarify how existing RPC estimators relate to one another, and provide a convenient reference for their properties at finite ensemble size.

In summary, an unbiased estimate of the predictable signal variance (equation \ref{equation:unbiased_var}) yields an ensemble-mean correlation, $r_{\infty}$, that is free from systematic ensemble-size bias for both reliable and unreliable forecasts. Inadmissible estimates of $r_{\infty}$ can be reported rather than clipped, as they indicate that the predictable signal cannot be resolved with the available ensemble size and number of cases. Applying the same signal variance estimate to the model-model correlation gives $\textnormal{RPC}_{\infty}$, an ensemble-size-invariant estimate of RPC that targets the same population quantity as $\textnormal{RPC}_{\textnormal{RV26}}$. In contrast, $\textnormal{RPC}_{\textnormal{E14}}$ and $\textnormal{RPC}_{\textnormal{H22}}$ are biased low by factors of $\gamma_N$ and $\sqrt{\gamma_N}$, respectively (equation \ref{equation:rpc_bias}), which depend only on the ensemble size and the model signal-to-noise ratio.

\newpage
\bibliographystyle{rss}
\bibliography{references}

\end{document}